\documentclass[ps,prl,twocolumn,showpacs,groupedaddress]{revtex4} % for review and submission
\usepackage{graphicx}
\usepackage{dcolumn}   % needed for some tables
\usepackage{bm}        % for math
\usepackage{amssymb}   % for math
\usepackage{amsmath}   % for math
\usepackage{epsfig}
\usepackage{enumerate}
\usepackage{natbib}

\begin{document}

\title{A Sparse SCF algorithm and its parallel implementation: Application to DFTB}

\author{Anthony Scemama$^1$, Nicolas Renon$^2$, Mathias Rapacioli$^1$}
\affiliation{
$^1$Laboratoire de Chimie et Physique Quantiques,
Universit\'e de Toulouse-CNRS-IRSAMC, France. \\
$^2$CALMIP,
Universit\'e de Toulouse-CNRS-INPT-INSA-UPS, France ; UMS 3667.
}

\date{\today}
\begin{abstract}
We present an algorithm and its parallel implementation for solving a self consistent 
problem as encountered in Hartree Fock or Density Functional Theory. 
The algorithm takes advantage of the sparsity of matrices through the use of
local molecular orbitals. The implementation allows to exploit efficiently
modern symmetric multiprocessing (SMP) computer architectures. As a first
application, the algorithm is used within the density functional based tight
binding method, for which most of the computational time is spent in the linear
algebra routines (diagonalization of the Fock/Kohn-Sham matrix). We show that with this algorithm 
(i)  single point calculations on very large systems (millions of atoms) can be performed on large SMP machines
(ii) calculations involving intermediate size systems (1~000--100~000 atoms) are also strongly accelerated and can run efficiently on standard servers
(iii) the error on the total energy due to the use of a cut-off in the molecular orbital coefficients can be controlled such that it remains smaller than the SCF convergence criterion.
\end{abstract}

\maketitle

\section{Introduction}

Density Functional Theory (DFT)\citep{hohenberg}, as routinely used with the practical formalism of Kohn and Sham,\citep{kohn} 
is a very efficient approach to access physical and chemical properties for a large amount of 
systems. Apart from the programs working with a grid to express the electron density, Kohn-Sham determinants are usually 
expressed on molecular orbitals (MOs) expanded on a basis set of $N$ atomic orbitals (AOs), and the electronic problem is solved iteratively until a self consistent 
solution is found. Technically, the na\"ive implementation involves the storage
of several matrices containing ${\cal O}(N^2)$ elements and the computation of several diagonalization steps, each with a ${\cal O}(N^3)$ computational complexity.
Hence, such an implementation of DFT can hardly address systems containing
more than several hundreds of atoms with current computational 
facilities. However, the physical intuition tells us that it is possible to
achieve linear scaling with the size of the system as the inter-particle interactions are essentially local.

In the past decades, there has been multiple developments of linear-scaling
techniques, reducing both the required memory and the computational
cost.
Those imply inevitably the use of sparse matrices: matrices transformed
such that almost all the information is packed into a small number of matrix elements 
(the non-zero elements).
The computational cost is also reduced since operations involving zeros are 
avoided.

The electronic problem in DFT consists in minimizing the electronic energy
with respect to the parameters of the density. These parameters can be expressed
whether as molecular orbital coefficients, or as density matrix elements.
The density matrix has the property to be local, so the density matrix is
naturally sparse, and this feature is exploited in density-matrix based
algorithms.
Density matrix schemes usually rely either on a polynomial expansion of the
density matrix including several purification
schemes,\citep{Palser,niklasson03,niklasson,mcweeny,challacombe,jordan,rubensson,rudberg,Rudberg10}
or on the direct minimization of the energy with gradient or conjugated gradient
schemes.\citep{Li93,Bowler,li2003,millam,challacombe,jordan,adhikari,daniels}
In these iterative approaches, a single step consists in performing a limited
number of sparse matrix multiplications. Then, the use of sparse linear
algebra libraries usually leads to linear scaling codes for very large systems.
Among these schemes, the matrix sign function or the orbital free
framework have recently reported linear scaling opening the
route to the calculation of millions of atoms at the DFT level.\citep{Hung2009163,cp2k} 
In schemes implying the expression of the MOs, the
linear scaling regime can be reached by first localizing the MOs in the
three-dimensional space and then using advanced diagonalization techniques,
usually in a divide and conquer framework.\citep{dixon,stewart}

The computational cost of the whole self consistent field (SCF) process can be seen as the sum of the
time spent in the building of the Fock/Kohn Sham matrix and the time spent in the
linear algebra, that is the multiple matrix diagonalizations. In this
work, we present a general algorithm to solve the linear algebra SCF problem
taking advantage of the sparsity of the systems and of the modern shared-memory
symmetric-multiprocessing (SMP) architectures. This algorithm is general and can
be in principle applied to any single-determinant based method (DFT, Hartree
Fock, and their semi-empirical counterparts). We have implemented this
algorithm in the deMon-Nano code\citep{deMonNano},
a density functional based tight binding (DFTB) program.
DFTB\citep{dftb1,dftb2,scc-dftb,frauenheim00,frauenheim02,augusto09} uses an approximate DFT scheme in which the elements of the Kohn Sham matrix
are computed efficiently from pre-calculated parameter tables. As in DFTB most
of the computational time is spent in the SCF linear algebra, it is a good
benchmark for the proposed implementation. In addition, DFTB is very well suited
to linear scaling schemes since it uses minimal and compact atomic basis sets
which are very favorable to have a very large fraction of zeros in the matrices.
Remark that DFTB has already been
improved in the literature by taking advantage of the sparsity of the matrices
to compute properties,\citep{DFTB+} combining the approach with the divide and conquer 
scheme\citep{Liu,frauenheim02} or with the matrix sign function scheme 
to achieve very large scale computations.\citep{cp2k} Recently Giese {\it et al} performed fast
calculations on huge water clusters describing intra-molecular interactions at
the DFTB level and inter-molecular interactions in a quantum mechanical
scheme.\citep{Giese:2013fk} 

Modern computers are very good at doing simple arithmetic operations (additions
and multiplications) in the CPU core, but are much less efficient at moving
data in the main memory. The efficiency of the data transfers is also not
constant: when the distance in between two data access increases, the
latency increases accordingly because of the multiple levels of cache and
because the data may be located on a memory module attached to another CPU
(cache-coherent Non-Uniform Memory Access, cc-NUMA architecture). Dense linear algebra can
exploit well such architectures because the computational complexity of a
matrix product or of a diagonalization is ${\cal O}(N^3)$, whereas the required
storage is ${\cal O}(N^2)$: the cost of the data movement becomes negligible
compared to the cost of the arithmetic operations and the processors are 
always fed with data quickly enough. In linear scaling methods both the
computational complexity and the data storage are ${\cal O}(N)$, and the
arithmetic intensity (the number of operation per loaded or stored byte) is
often not high enough to have CPU-bound implementations. These algorithms are
then memory-bound. In addition, the data access patterns are also not as
regular as in the dense matrix algorithms where all the elements of the arrays
can be explored contiguously. In that case, the hardware mechanisms that prefetch
data into the caches to hide the large memory latencies will not work as
efficiently, and these algorithms become {\em memory-latency-bound}. Moreover,
the irregular access patterns may disable the possibility of using vector
instructions, reducing the performance of arithmetic operations and data
movement. As the efficiency of memory accesses decreases with the size of the
data, the wall time curves are not expected to be linear in ``linear-scaling''
implementations. Linear wall time curves appear in two situations.
The first situation is when the data access is always in the worst condition
(with the largest possible latency). The second
situation is when the cost of arithmetic operations is sufficiently high to
dominate the wall time. This happens when the arithmetic intensity is high (the
same data can be re-used for a large number of operations), or when the
involved arithmetic operations are expensive (trigonometric functions,
exponentials, divisions, impossibility to use vector instructions, {\it etc}).
Therefore, implementing an efficient linear scaling algorithm requires to
carefully investigate the data structures to optimize as much as
possible the data transfers. The sole reduction of the number of arithmetic
operations is not at all a guarantee to obtain a faster program: doing some
additional operations can accelerate the program if it can cure a bad data
access problem. A final remark is that linear scaling is {\em asymptotic}. In
practice, we make simulations in a finite range, and it makes sense to
find the fastest possible implementation in this particular range. For
instance, if an ${\cal O}(N^2)$ algorithm is able to exploit the hardware 
such that the ${\cal O}(N)$ and ${\cal O}(N^2)$ curves intersect in a domain
where $N$ is much larger than the practical range, the quadratic algorithm is
preferable.
So in this work, we did not focus on obtaining asymptotic linear wall time
curves but we tried to obtain the fastest possible implementation for systems
in the range of $10^3$--$10^6$ atoms with controlled approximations.

After presenting the general algorithm, we give the technical implementation
details that make the computations efficient for each step of the algorithm.
Throughout the paper, the implementation will be illustrated
by a benchmark composed of a set of boxes with an increasing number of water
molecules.

\section{Algorithm and computational details}

The general algorithm consists in the solving the self-consistent Roothaan equations by~:
\begin{enumerate}[I)]
\item  \label{step_kmeans} Defining neighbouring lists of atoms and MOs 
\item  \label{step_guess} Proposing an initial guess of local MOs
\item  \label{step_ortho} Othornormalizing the MOs
\item  \label{step_fock} Computing the electronic density and the Fock/Kohn Sham matrix
\item  \label{step_diag} Partially diagonalizing the Fock/Kohn Sham matrix in the MO basis
\end{enumerate}
Steps \ref{step_fock}~and~\ref{step_diag} are iterated to obtain the
self-consistent solution. Note that this algorithm is general: only
step~\ref{step_fock} is specific of the underlying
method (Hartree-Fock, DFT, and their semi-empirical frameworks). In the following we define
${\bf H}$ as the Fock/Kohn Sham operator expressed in the atomic basis set and
${\bf S}$ the overlap matrix of the atomic basis functions.
Due to the strongly sequential character of the algorithm, each step was parallelized
independently using OpenMP\citep{openmp} in an experimental version of the
deMonNano package.\citep{deMonNano}

In recent years the number of compute nodes in x86 clusters has been roughly
constant with an increasing number of CPU cores per node. In addition, the
number of cores increases faster than the available memory, so the memory per
CPU core decreases. This evolution of super-computers is not well suited
to implementations handling parallelism only with the Message Passing Interface (MPI)\citep{mpi} 
where each CPU core is running one MPI process.
First, as the number of CPU cores per node increases the number of MPI
processes per node increases correspondingly, and the total required memory per node increases
proportionally. Secondly all the MPI processes running on one node will share the same
network interfaces, and the bottleneck on the network
will become more and more important, especially in collective communications.
Hybrid parallel implementations usually combine distributed processes with
threaded parallelism. The most popular framework is the combination of MPI for distributed parallelism and
OpenMP for threaded parallelism.
Implementations running one distributed process per node or socket where each process uses multiple threads
are expected to be much more long-lasting.
In this context, we chose to
use a shared-memory approach for the calculation of the energy, and leave
the distributed aspect for coarser-grained parallelism which will be
investigated in a future work. Indeed, potential energy surfaces of large
systems involve multiple local minima due to the large number of atomic degrees of
freedom, and the chemical study of large systems implies the resolution
of the Schr\"odinger equation at many different molecular geometries (geometry
optimization, molecular dynamics, Monte Carlo, {\it etc}) that may be distributed
with a very low network communication overhead.

The systems chosen as a benchmark are cubic boxes of liquid water, containing 
from 184 to 504 896 molecules (from 552 to 1.5 million atoms).
Running the benchmark consists in computing the single-point energy
of each independent box at the DFTB level.
All the reported timings correspond to the {\em wall time}.
The calculations were performed on two different architectures. The first one
is a standard dual socket Intel Xeon E5-2670 (each socket has 8 cores, 20~MiB Cache,
2.60~GHz, 8.00~GT/s QPI) with Hyperthreading and turbo activated: CPU frequencies
were 3.3~Ghz for 1--4 threads, 3.2~GHz for 8 threads, and 3.0~GHz for 16--32
threads.
The second architecture is an SGI\textregistered{ }Altix UV 100 with 48 sockets (384 cores)
and 3~TiB of memory. As the Altix UV 100 is based on the so-called ``ccNUMA''
architecture, in our case 24 dual-socket blades are interconnected thanks to the
SGI NUMAlink\textregistered{ }technology with a 2D-Torus topology. Hence this Single System
Image (SSI) allows to access to the 3~TiB of memory as one
single unified memory space. Each blade of the Altix machine is equipped with
128~GiB RAM and two Intel Xeon E7-8837 sockets (each one with 8 cores, 24~MiB Cache, 2,67~Ghz, 6,4~GT/s QPI).
The Hyperthreading and Turbo features were disabled.
The memory latencies measured on such a system are higher than the 80~ns measured on the dual-socket server:
\begin{itemize}
\item 195~ns when the memory module is directly attached to the socket
\item 228~ns when accessing memory attached to the other socket of the blade through the Quick Path Interconnect link
\item 570, 670, 760, 875, 957~ns when accessing directly memory located in another blade through the NUMAlink\textregistered{ }interconnect
(larger latencies correspond to an increasing number of jumps in the NUMAlink\textregistered{ }network before the target blade can be reached).
\end{itemize}

On the dual-socket architecture, the code was compiled with the Intel Fortran
Compiler version 12.1.0 with the following options: {\tt -openmp -xAVX
-opt-prefetch=4 -ftz -ip -mcmodel=large -pc 64}. The threads were pinned to
CPU cores using the {\tt taskset} tool. For jobs using less than 16 cores, the
threads were scattered among the sockets: for instance, a 4 thread job was
running 2 threads on each socket.

On the Altix-UV, the code was compiled with the Intel Fortran Compiler version
12.0.4 with the following options: {\tt -openmp -g -xSSE4.2 -opt-prefetch=4
-ftz -ip -mcmodel=large -pc 64}. The threads were pinned to CPU cores using
the {\tt omplace} (SGI MPT\textregistered) tool and runs were always performed such that the blades were
fully active (16 threads running of 16 cores for each used blade).

\section{Implementation}

\subsection{Sparse storage}

To make calculations feasible for large systems it is mandatory to reduce the
storage of the matrices. Obviously, we use a sparse representation of the
matrices by storing only the matrix elements with an absolute value above a
given threshold.

Many different sparse storage schemes exist. The choice of the format depends
on the work that has to be done with the data. For instance the compressed
sparse row and the compressed sparse column formats,\citep{Kincaid}
as well as their blocked variants are usually chosen when using a sparse solver
as a black box. Compressed Diagonal Storage format is mostly used in the finite-element
community since it is well suited to diagonal band matrices,\citep{melhem1987toward}
and the Skyline Storage format is usually chosen when no pivoting is
required for a LU factorization.\citep{duff1986direct}

The algorithm detailed here implies many rotations between vectors appearing in
an unpredictable order. Knowing this, it is convenient to sacrifice some
memory and use a scheme where the sparse vectors are equally spaced using
padding, such that their size can be easily expanded or reduced after a
rotation. Therefore, we use the List of Lists (LIL) format implemented as
two-dimensional arrays where the first dimensions of the arrays has a fixed
size $s_m$, a little larger than an estimation of the largest encountered size
in the calculations, which should be constant for large systems.

\begin{figure}
 \begin{center}
  \includegraphics[width=0.7\columnwidth]{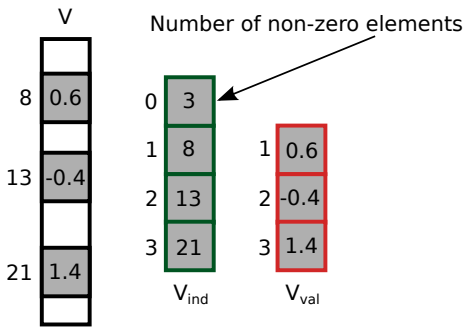}
 \end{center}
 \caption{Sparse representation of a vector in the LIL format.}
 \label{fig:sparse_vector}
\end{figure}

Using Fortran notation, a sparse array {\tt A(N,N)} is represented by an
integer array of indices {\tt A\_ind(0:sm,N)} and a real or double-precision
array of values {\tt A\_val(s\_m,N)}. The access to a column is direct using
the second index. A column is stored as the list of non-zero values in {\tt
A\_val} and the corresponding row indices in {\tt A\_ind} (see
figure~\ref{fig:sparse_vector}). The elements at {\tt A\_ind(0,:)} contain the
number of non-zero elements in the corresponding columns.

Using compiler directives, the floating-point arrays are aligned on a 512-byte
boundary, and $s_m$ is constrained to be a multiple of 8 elements. In this way,
all the columns of the array are properly aligned on the 256-bit boundary
allowing aligned vector load and store instructions using both the SSE or the
AVX instruction sets. Hence, each column starts whether at the beginning of a
cache line, or at the middle of a cache line (in single precision).

If $s_m$ is set to a multiple of 512 double-precision elements, the columns are
spaced by a multiple of 4~KiB. In this case, there will be conflict in the cache when
loading two columns since they will have the same address in the cache. To avoid this
so-called {\em 4k-aliasing}, we constrain $s_m$ to follow the rule
\begin{equation}
 s_m = \alpha \times 512 + 24 %\,, \alpha \le 9
\end{equation}

For the largest system presented here, a box made of 504~896 water molecules
(1.5 million atoms), the maximum value of the total resident memory was 982~GiB,
which corresponds to an average of $\sim$ 680~KiB per atom.

\subsection{Defining neighbouring lists of atoms and MOs (step~\ref{step_kmeans}) }

Two levels of localization or neighbouring are used in this work. 
Two atoms are always considered as neighbours, unless all the off-diagonal elements of the 
overlap matrix ${\bf S}$ and of the Fock/Kohn Sham matrix ${\bf H}$ involving one AO centered on each atom
are zero. At the beginning of the program, 
an atomic neighbouring map is built and is used to avoid computing zero elements 
of ${\bf S}$ and ${\bf H}$ in the AO basis.

It is also convenient to use groups of localized MOs. 
This is done by generating an initial guess of MOs (see below)
with non-zero coefficients on a limited number of AOs centered on atoms in the
same region of space.
During the orthonormalization and SCF process, MOs belonging to one group can of course delocalize and have coefficients on AOs
centered on atoms belonging to another group. We however keep the initial group attribution of MOs during all the SCF process as
the MOs are not expected to delocalize sufficiently to change the MO group attribution.
Two groups are considered as neighbours if there exists a distance between one atom of the first group and an atom of the second one which is smaller than a given value (in practice we use 20~bohr).
A map of MO group neighbours is built and we will assume that, in the MO basis, blocks of ${\bf H}$ and ${\bf S}$ involving 
non-neighbouring groups will always be zero.
This assumption relies on the fact that as both occupied and virtual MOs are
localized they are not expected to delocalize by more than 20~bohr during the
SCF process. This assumption can be checked after the SCF has converged, and
has not to be misinterpreted as a screening of long range $1/R$ interactions.

We now describe the procedure to build MO groups. 
Atomic orbitals centered on atoms that are geometrically far from
each other have a zero overlap, and don't interact through the
Fock/Kohn Sham operator. All the atoms of the system are reordered such that
atoms close to each other in the three-dimensional space are also
close in the list of atoms. To achieve this goal, we use a
constrained version of the {\em k-means} clustering
algorithm.\citep{kmeans}

A set of $m$ centers ({\em means}) is first distributed evenly in the
three-dimensional space. Every atom is linked to its closest center
with the constraint that a center must be linked to no more than
$N_{\rm link}$ atoms~: If an atom can't be linked to a center because
the latter already has $N_{\rm link}$ linked atoms, the next
closest center is chosen until an available center is found. 
Each center is then moved to the centroid of its connected atoms and
the procedure is iterated until the average displacement of the
centers is below a given threshold, or until a maximum number of
iterations is reached (typically 100). Finally, the new list of atoms
is built as the concatenation of the lists of atoms linked to each
center.

Because of the constraint to have no more than $N_{\rm link}$ atoms attached to
one center, the atom-to-center attribution depends on the order in which the
atom list is explored and multiple solutions exist. This has no effect in the
final energy since the definition of the neighbours depends on the atom-atom
distances. However, it can have an impact on the computational
time as the lists of neighbours can have different lengths. To cure
this problem, the list of centers is randomly shuffled at every
k-means iteration.

Each k-means attribution step is parallelized over the atoms.
The $N_{\rm link}$ constraint introduces a coupling between the
threads which is handled using an array of OpenMP locks: a lock
is associated with each center, and the lock is taken by the thread before
modifying its list of atoms and then released. The displacement of the centers
is trivially parallelized over the centers, and this
initialization step takes 2--5~\% of the total wall time.

\subsection{Initial guess of molecular orbitals (step~\ref{step_guess})}

For each molecule of the system, a non-self-consistent
calculation is performed. The molecular orbitals (MO) matrix ${\bf C}$
in the atomic orbitals (AO) basis set is built from the concatenation 
of all the lists of occupied MOs of each group and all the lists
of virtual MOs of each group.

These MOs are not orthonormal but have the following properties: 
\begin{itemize}
\item MOs are local: they have non-zero coefficients only on the
basis functions that belong to the same molecule
\item MOs belonging to the same molecule are orthonormal
\item MOs $(i,j)$ belonging to MO groups that are not
neighbours have a zero overlap and don't interact: $({\bf C^\dagger S
C})_{ij} = ({\bf C^\dagger H C})_{ij} = 0$
\end{itemize}
If the MOs don't delocalize by more than 20~bohr, which was the case on all our benchmarks,
the large off-diagonal blocks of zeros corresponding to the last
point persist in the orthonormalization and SCF processes, so ${\bf C}$
always stays sparse.

The non-SCC calculations are parallelized over the molecules using
a dynamic scheduling to keep a good load-balancing, and this initialization
step takes less than 1\% of the total wall time.

\subsection{Orthonormalization of the molecular orbitals (step \ref{step_ortho})}

The MOs are orthonormalized by diagonalizing iteratively the
${\bf C^\dagger S C}$ matrix. At this point, ${\bf C}$ is already stored in a sparse format.
\begin{enumerate}[a.]
 \item \label{comp_s} ${\bf S}$ is computed (or read from memory) only in elements involving AOs of atomic neighbours
 \item \label{csc} ${\bf C^\dagger S C}$ is computed
 \item \label{norm} The MOs are normalized using diagonal elements of ${\bf
C^\dagger S C}$
 \item \label{combi} Combinations (approximate rotations) are performed between the pairs of MOs
 $(i,j)$ which involve the largest off-diagonal elements of the MO overlap matrix:
  \begin{eqnarray}
   C'_{ki} &=& C_{ki} - \frac{C_{kj} ({\bf
   C^\dagger SC})_{ij}}{2} \nonumber \\
   C'_{kj} &=& C_{kj} - \frac{C_{ki} ({\bf
   C^\dagger SC})_{ij}}{2}  \nonumber 
  \end{eqnarray}
 \item \label{csc_loop} Return to step 3 until the largest off-diagonal element is
 below a given threshold
\end{enumerate}

Note that, if the computational cost of building of the overlap matrix elements
is comparable to reading it from the memory, it is suitable to compute the
overlap matrix elements only when they are needed in the computation of ${\bf
C^\dagger S C}$ to reduce the amount of storage.

The parallelization of the MO combinations (step~\ref{combi}) is
not trivial. If one combination of MOs $(i,j)$ is being performed by one
thread, any combination of pairs of MOs involving either $i$ or $j$ can't be
performed simultaneously by another thread. 
First, we prepare an array of OpenMP locks, one lock for each MO. Then, we
dress the list of orbital pairs $(i,j)$ to combine, those to which corresponds
an off-diagonal element $|{\bf C^\dagger S C}|_{ij} \ge 10^{-3} \eta$, where
$\eta$ is the largest off-diagonal element of $|{\bf C^\dagger S C}|$. The
total length of the list is $N_{\rm pair}$. Each thread executes simultaneously
the following steps:
\begin{enumerate}
\item \label{s1} Pick the first combination $(i,j)$ and go to step~\ref{s3}
\item \label{s2} Pick the next combination $(i,j)$
\item \label{s3} If the combination $(i,j)$ is already done, go to step~\ref{s2}
\item \label{s4} Try to take lock $L_i$ with {\tt OMP\_TEST\_LOCK}. If not possible,
  go to step~\ref{s2}
\item \label{s5} Try to take lock $L_j$ with {\tt OMP\_TEST\_LOCK}. If not possible,
  free lock $L_i$ and go to step~\ref{s2}
\item \label{s6} Combine $i$ and $j$ and mark $(i,j)$ as done
\item \label{s7} Free locks $L_i$ and $L_j$
\item \label{s8} Go back to step~\ref{s2} until the end of the list is reached
\item \label{s9} Go back to step~\ref{s1} until all combinations are done
\end{enumerate}
The list of combinations is stored in an array, which can be large. Using the
algorithm as presented explores many times the long list of combinations. After
the first pass most of the combinations are done and looping over the whole
array will spend a lot of time checking that a combination has already been
done. An optimized approach consists in looping over only the first half of the array ($k
\in [1,N_{\rm pair}/2]$). After a combination has been done, the elements $k$ and
$k+N_{\rm pair}/2$ of the list are swapped. In this way, when the $N_{\rm
pair}/2$ first elements have been explored, the first $N_{\rm pair}/2$ elements
of the list are combinations to be done. Then, $N_{\rm pair}$ is set to $N_{\rm pair}/2$,
and the procedure is repeated until $N_{\rm pair} < N_{\rm basis}$. Using this
strategy allows to loop over only combinations to do and it avoids to check if
a combination has already been done (loop over steps~\ref{s2} and~\ref{s3}). At
this point, there remains a few combinations to perform scattered in the array.
Those last combinations are
handled with the algorithm presented above (steps \ref{s1}--\ref{s9}), but using blocking
functions with {\tt OMP\_SET\_LOCK} instead of the non-blocking
{\tt OMP\_TEST\_LOCK}. As most of the combinations have already been done,
the probability of trying to lock simultaneously the same MO will be low
and the locking steps will be occasional. Therefore, using blocking locks
is relevant here since it will allow to perform only one traversal of the array
to complete the remaining combinations.

The orthonormalization takes 40--50\% of the total wall time. Half of the time is
spent in the MO rotations and the other half is spent in 
the computation of ${\bf C^\dagger S C}$ which is discussed in detail below. 

\subsection{Partial diagonalization of the Fock/Kohn Sham matrix (step~\ref{step_fock})}

The wave function is invariant with respect to unitary transformations among
doubly occupied orbitals or among virtual orbitals. Therefore, the full
diagonalization of ${\bf H}$ is not necessary to obtain the SCF minimum 
energy and properties. It is sufficient to annihilate only the matrix elements of the Fock/Kohn Sham matrix 
expressed in the MO basis involving one occupied and one virtual orbital. To do that, we use an approach
similar to the orthonormalization procedure: we apply Jacobi rotations
between in the occupied and virtual orbitals, as proposed by Stewart.\citep{stewart}
To preserve the orthonormality of the MOs, the orbitals are combined using an
accurate rotation but with an approximate angle.

\begin{enumerate}
 \item Compute ${\bf H}$ only in elements involving AOs of neighbouring atoms
 \item Compute ${\bf C^\dagger H C}$ as explained in next section
 \item \label{ss3} Rotate the pair of MOs $(i,j)$ which involves the largest off-diagonal element:
  \begin{eqnarray}
    p &=& \frac{({\bf C^\dagger HC})_{ii} - ({\bf C^\dagger HC})_{jj} }{2  ({\bf C^\dagger HC})_{ij}} \nonumber \\
    t &=& \left( p+ \operatorname{sgn}(p) \sqrt{1+p^2} \right)^{-1} \nonumber  \\
    c &=& \left( 1+t^2 \right)^{-1/2} \nonumber \\
    s &=& tc \nonumber \\
   C'_{ki} &=&  c C_{ki} + s C_{kj} \nonumber \\
   C'_{kj} &=& -s C_{kj} + c C_{ki} \nonumber 
  \end{eqnarray}
 \item Return to step~\ref{ss3} until the largest off-diagonal element is below
a given threshold or if the number of rotations is equal to 10 times the
number of occupied MOs.
\end{enumerate}

As mentioned for the orthonormalization, it can be suitable to compute the
elements of the Fock/Kohn Sham matrix directly when needed at the second step
to avoid the storage of a matrix.

\begin{figure}
 \centering
 \includegraphics[width=0.75\columnwidth]{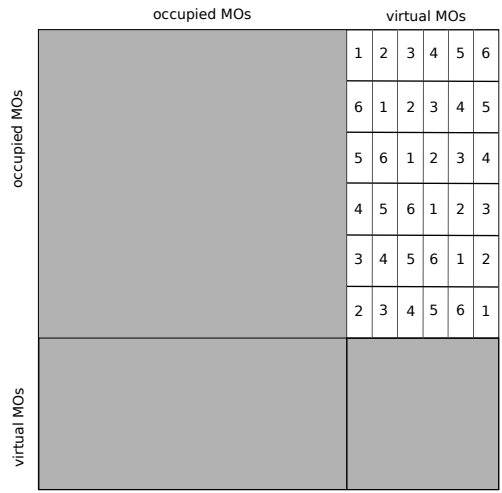}
 \caption{Representation of the ${\bf C^\dagger H C}$ matrix. Only one off-diagonal
  block is annihilated by pseudo-Jacobi rotations. The numbers identify iterations.
  On the first iteration, blocks labeled by 1 are annihilated in parallel, {\it etc}}
 \label{fig:jacobi}
\end{figure}

Avoiding to perform occupied-occupied and virtual-virtual rotations keeps both
the occupied and virtual MOs localized during the optimization.\citep{angeli,maynau}
In this way, the ${\bf C}$ matrix stays sparse.
The ${\bf C^\dagger H C}$ matrix is displayed in figure~\ref{fig:jacobi}. As
the matrix is symmetric and only the occupied-virtual rotations are considered,
only one occupied-virtual block is computed for the orbital rotations,
appearing in white on the figure. This block can be divided in sub-blocks using the MO groups definition.

The parallelization of the rotations is
implemented such that one orbital can never be rotated simultaneously by two
threads~: each thread performs a rotation in a block of the occupied-virtual
elements. The figure shows the outer loop iterations. On the first iteration, the
threads perform rotations in all the blocks labeled by 1. After a synchronization,
on the second iteration they perform rotations in the blocks labeled by 2, {\it etc}.
Blocks between non neighbouring MO groups are not considered for the rotation process.
Let us mention that in practice the occupied-virtual ${\bf C^\dagger H C}$ block is computed 
and stored in a sparse format.

The time spent in the MO rotations represents 3--6\% of the total wall time.
The bottleneck is the calculation of ${\bf C^\dagger H C}$.
Adding the timing of ${\bf C^\dagger H C}$ yields 33--40\% of the total wall time.

\subsection{Computation kernel for the matrix products ${\bf C}^\dagger {\bf X C}$ (steps~\ref{step_ortho} and~\ref{step_diag})}
\label{CXC}

From the two previous sections, it appears that the main hot spot 
is the computation of double sparse matrix products of the form
${\bf C}^\dagger {\bf X C}$ where ${\bf X}={\bf S}$ for the orthonormalization 
part and ${\bf X}={\bf H}$ for the diagonalization part. We explain here how this 
step is performed taking advantage of the sparsity of matrices ${\bf C}$, ${\bf H}$
and ${\bf S}$. 

In another work, an efficient dense$\times$sparse matrix product
subroutine for small matrices was implemented in the QMC=Chem quantum
Monte Carlo software.\citep{qmcchem} These matrix products are
hand-tuned for Intel Sandy-Bridge and Ivy-Bridge micro-architectures,
and have reached more than 60\% of the peak performance of a CPU core.
To obtain such results, some programming constraints are necessary
and are detailed in the next section.

In what follows, large sparse$\times$sparse matrix products are
performed as a collection of small dense$\times$sparse products
avoiding the zero-blocks connecting non-neighbouring MO groups.

\subsubsection{Small dense$\times$sparse matrix products}

\begin{figure}
 \begin{center}
  \includegraphics[width=0.9\columnwidth]{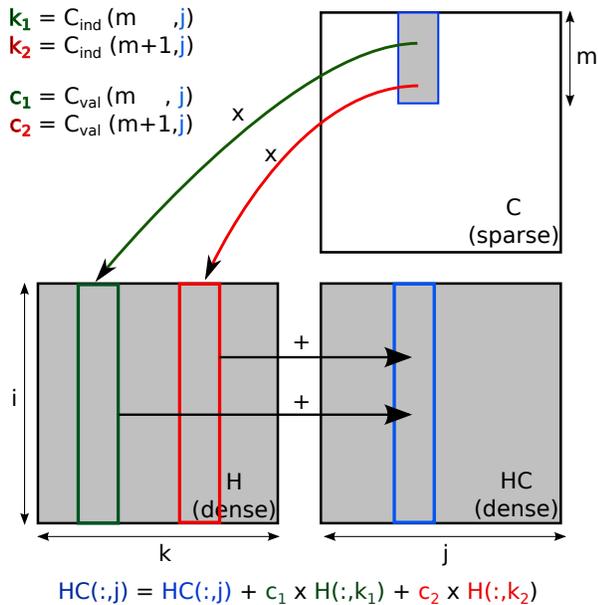}
 \end{center}
 \caption{Dense$\times$sparse matrix product.}
 \label{fig:sparse_prod}
\end{figure}

Let us recall the dense$\times$sparse matrix products of
Ref~\citep{qmcchem}.
In the multiplication routine, the loops are reordered such that
vectorization is possible, giving a dense matrix in output (see
figure~\ref{fig:sparse_prod}).
As the innermost loop has a small number of iterations,
the following constraints are mandatory to obtain performance:
\begin{itemize}
 \item All arrays have to be 32-byte aligned using compiler
directives
 \item The leading dimensions of arrays should be multiples of 8
elements such that all the columns are 32-byte aligned
\end{itemize}
Without these constraints, the compiler will generate both the scalar
and the vector version of the loop, and the scalar version will be
used as long as the 32-byte alignment is not reached (peeling loop) or
if a register of 4 double precision or 8 single precision elements
cannot be filled (tail loop). In such a case, a 16-iteration loop in single precision
will perform 8 scalar loop cycles and one vector loop cycle.
Using the alignment and padding constraints, the compiler
can produce 100\% vector instructions with no branching and will
execute 2 vector loop cycles unrolled by the compiler, resulting in a speedup of more
than 5. 
Note that these constraints contribute to the choice of the memory layout
for sparse matrices explained previously.

\subsubsection{Computation of ${\bf C^\dagger X C}$}

For each thread, the algorithm is the following:
\begin{enumerate}
 \item \label{xc} Compute 16 contiguous columns of ${\bf XC}$ in a dense representation
 \item \label{transpose} Transpose the result to obtain 16 rows of ${\bf C^\dagger X}$
 \item \label{mul_c} Multiply on the right by ${\bf C}$ to obtain 16 rows of ${\bf C^\dagger X C}$
 \item \label{sparsify} As ${\bf C^\dagger X C}$ is symmetric, transpose the
   dense 16 rows of ${\bf C^\dagger X C}$ into a sparse representation of 16
   columns of ${\bf C^\dagger X C}$
 \item \label{iterate} Iterate until all the columns of ${\bf C^\dagger X C}$ are built
\end{enumerate}

In step~\ref{xc}, dense$\times$sparse matrix products are performed using
only the $16\times16$ blocks of ${\bf X}$ containing at least one non-zero
element. This exploits the sparse character of ${\bf X}$ by avoiding large
blocks of zeros using the atomic neighbour list.
The $16\times16$ shape of the arrays is constant and known at compile time,
and all the columns of the $16\times16$ arrays are 32-byte aligned.
This permits the compiler to remove the peeling and tail loops
and generate a fully vectorized code.

The 16 dense columns of ${\bf XC}$ are not stored as a ($N_{\rm
basis}$,16)-array, but are instead represented as an array of $16\times 16$ matrices:
the dimension of the array is $(16,16,\lfloor N_{\rm basis}/16\rfloor+1)$. The reason
for this layout is twofold. First, the result of ${\bf XC}$ is
written linearly into memory. The spatial locality in the cache is improved and this
reduces the traffic between the caches and the main memory.
Secondly, step~\ref{transpose} can be performed very fast since the
transposition of ${\bf XC}$ consists in $\lfloor N_{\rm
basis}/16 \rfloor +1$ in-place transpositions of contiguous $16\times 16$ matrices.
Using the ($N_{\rm basis}$,16) layout, the transposition would
involve memory accesses distant by $N_{\rm basis}$ elements suffering
from a high memory latency overhead for large systems since
the hardware prefetchers prefetch data only up to a memory page boundary.
Using our layout, the transposition is always memory-bandwidth bound as
each $16\times 16$ transposition occurs in the L1 cache and the next $16\times 16$ matrix
is automatically prefetched.

In step~\ref{mul_c}, the dense$\times$sparse product is performed using only
the columns of ${\bf C}$ belonging to MO groups which are neighbours of the MO groups 
to which the 16 rows of ${\bf C^\dagger X}$ belong. Again, the loop
count of the inner-most loop is 16 and this part is fully vectorized
as all the columns of the arrays are properly aligned.

\subsubsection{Parallel efficiency}

\begin{figure}
 \centering
 \includegraphics[angle=270,width=\columnwidth]{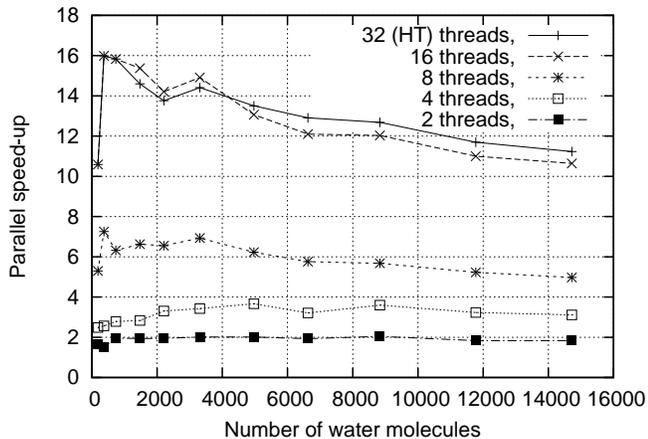}
 \caption{Parallel speed-up of ${\bf C^\dagger X C}$ as a function of the number of
water molecules obtained on the dual-socket server.
The speed-ups were renormalized to remove the effect of the turbo feature.
The 32 threads results were obtained using Hyperthreading (2 threads per physical core).}
 \label{fig:cxc_speedup_sv7}
\end{figure}

\begin{figure}
 \centering
 \includegraphics[angle=270,width=\columnwidth]{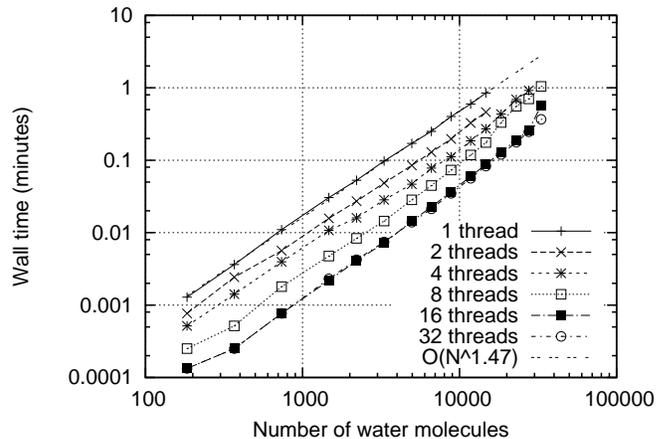}
 \caption{Wall time of ${\bf C^\dagger X C}$ as a function of the number of
water molecules obtained on the dual-socket server.
The 32 threads results were obtained using Hyperthreading (2 threads per core).}
 \label{fig:cxc_scaling_sv7}
\end{figure}

The parallel speedup curve of the ${\bf C^\dagger X C}$ products measured on
the dual-socket server is shown in figure~\ref{fig:cxc_speedup_sv7}.
For small sizes, the speed-up is nearly optimal (the perfect value of 16 obtained here
is fortunate, since measuring very small wall-times contains large uncertainties).
This can be explained by the fact that most of the matrices can fit in the level 3
cache. Then, as the size of the {\bf C} matrix increases, more and more pressure is
made between the level 3 cache and the memory, as every core has to read the 
whole ${\bf C}$ matrix when multiplying the 16-rows with ${\bf C}$.
The latency of data access to {\bf C} increases accordingly and the processor cores
can't be kept busy. This is confirmed by the fact that Hyperthreading improves
the parallel efficiency by giving some work to a core for one thread while
another thread is stalled. The speed-up converges to a limit of 11 for a
16-core machine. The wall time curve of figure \ref{fig:cxc_scaling_sv7} shows
a scaling with the number of atoms of ${\cal O}(N^{1.47})$. The scaling is not
linear, because of the non-uniformity of the memory latencies.

\subsection{Specific implementation for DFTB }
\label{DFTB}
We now turn to the specific implementation of the previously described algorithm for
density functional based tight binding (DFTB), an approximate DFT scheme.\citep{dftb1,dftb2,scc-dftb,elstner2006scc,frauenheim00,frauenheim02,augusto09}
The electronic problem is only solved for valence electrons and
molecular orbitals are expressed in a minimal atomic basis set. 
The DFTB Kohn-Sham (KS) matrix is: 
\begin{equation}
\label{H_dftb}
H_{\mu\nu}=H^0_{\mu\nu}+\frac{1}{2}S_{\mu\nu}\sum_\xi (\gamma_{\alpha\xi}+\gamma_{\beta\xi})\Delta q_\xi 
\end{equation}
$ H^{0}_{\mu \nu} $ is the KS matrix element between orbitals $\mu$ of atom $\alpha$ and $\nu$ 
of atom $\beta$ at
 a reference density. This value is usually interpolated from pre-calculated DFT curves (the so-called Slater Koster
tables\citep{SlaterKoster54}). One-site $\gamma _{\alpha\alpha}$ are obtained from Hubbard parameters and inter-atomic
$\gamma_{\alpha \xi}$ 
depend on the distances between atoms $\alpha$ and $\xi$, and contain the long range $1/R$ Coulomb
interaction. 
$\Delta q_\xi=q_\xi-q^0_\xi$ is the charge fluctuation of atom $\xi$ calculated
with the M\"ulliken analysis scheme: 
\begin{equation}
\label{pop}
 q_\xi=\sum_{i}^{N_{\rm MOs}} n_i \sum_{\nu}^{N_{\rm AOs}} \sum_{\mu \in \xi} C_{i\mu}C_{i \nu} S_{\mu\nu} = \sum_{\nu}^{N_{\rm AOs}} \sum_{\mu \in \xi} P_{\mu \nu} S_{\mu\nu}
\end{equation}
where ${\bf P}$ is the density matrix in the AO basis set. 

Computing the atomic M\"ulliken population of Eq.~\ref{pop} could be an expensive
part that has to be efficiently parallelized. The parallelization can be done
either on the MOs (one thread deals with a subset of MOs and computes their
contributions to the atomic populations of the whole system) or on the atomic
population (one thread computes the atomic population of some atoms arising
from all the MOs). The first scheme is limited by storing data whereas the
second one is limited by reading data. Storing data in memory is more costly
than reading data, so the second scheme was chosen.
An acceleration can be achieved considering the spatial localization of molecular 
orbitals (only a limited number of MOs contribute to the population of a given atom).
For each atom $\alpha$ we dress the list of the indexes of MOs having a
non-zero contribution on atom $\alpha$. Then, we compute a set of rows of the
density matrix $P_{\mu \in \alpha, \nu} =\sum_i C_{i\mu} C_{i\nu}$, where $\nu$
is such that $S_{\mu,\nu}\neq 0$ (using the atomic neighbours list),
and the sum over MOs $i$ only runs on the MOs listed as contributors
on atom $\alpha$. $q_{\xi}$ is then computed from equation Eq.~\ref{pop}.
The calculation of the charges $q_\xi$ takes typically 3\% of the total wall time,
and the observed scaling is ${\cal O}(N^{1.1})$. Note that we limit the number
of threads to a maximum of 16~in this part. Indeed, this section is extremely
sensitive to memory latency as the arithmetic intensity is very low.
Using more than 16 threads would need inter-blade communications on the Altix
UV which always hurt the performance for this particular section.

From the computational point of view, it is better to compute first
a vector $Q_\alpha=\sum_\xi \gamma_{\alpha\xi} \Delta q_\xi $ to rewrite Eq.~\ref{H_dftb} as: 
\begin{equation}
\label{H_dftb2}
H_{\mu\nu}=H^0_{\mu\nu}+\frac{1}{2}S_{\mu\nu} (Q_\alpha+Q_\beta)
\end{equation}
The computation of the $Q_\alpha$ values requires ${\cal O}(N^2)$ operations.
This term contains actually the long range $1/R$ contributions to the Coulomb
energy and a linear scaling on this part could only be reached at the price of
approximations.  We decided not to screen these long-range contributions in
order to keep the same results as those one would obtain from a standard
diagonalization scheme, the computational cost remaining acceptable (less than
20\% of the total wall time for the largest systems of our benchmarks).

\section{Results obtained on the whole scheme}

\subsection{Sparsity and numerical accuracy}

In the deMon-Nano program, the convergence criterion of the SCF procedure is
based on the largest fluctuation of the M\"ulliken charge~:
\begin{equation}
 \delta q_{\max} = \max_i  |q_i^{(n)} - q_i^{(n-1)}|
\end{equation}
where $q_i^{(n)}$ denotes the M\"ulliken charge of atom $i$ at SCF iteration $n$.
As this criterion is local, it ensures that all the different regions of the system
have converged to a desired quality, as opposed to a criterion based on the total energy
of the system which is global.
We define the {\em sparse threshold} $\epsilon$ as a quantity used to adjust the
sparsity of the MO matrices. $\epsilon$ is set ten times lower than the target SCF
convergence criterion $\delta q_{\max}$, such that if more precision is
requested for the SCF, the matrices become less and less sparse. In the limit
where the requested SCF convergence threshold is zero, the matrices are dense
and the exact result is recovered.

Once the initial guess is done, the initial MOs are improved by running an SCF with very loose
thresholds: $\epsilon \le 10^{-3}$ and $\delta q_{\max} \le 10^{-1}$ a.u. 
The calculation is then started using the thresholds given by the user in input. In what
follows, the thresholds were $\epsilon \le 10^{-6}$ and $\delta q_{\max} \le 10^{-5}$ a.u.
Typically 8 SCF iterations were needed to reach a convergence of $\delta
q_{\max} \le 10^{-1}$ a.u., and 9--13 additional iterations were necessary for
$\delta q_{\max} \le 10^{-5}$ a.u.

\begin{figure}
 \centering
 \includegraphics[angle=270,width=\columnwidth]{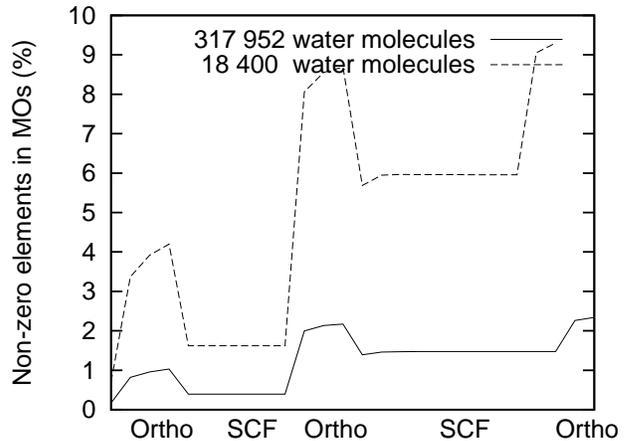}
 \caption{Percentage of non-zero elements during the execution of the program.}
 \label{fig:sparsity}
\end{figure}

The orthonormalization needs to be very accurate, so all the orthonormalization
procedure is run in double precision. After each rotation of pairs of MOs,
the MO coefficients ${\bf C}_{ki}$ such that $|{\bf C}_{ki}| < 10^{-3}
\epsilon$ are set to zero. This cut-off is chosen much smaller than $\epsilon$
to ensure that the quality of the orthonormalization will be sufficient.
However, this leads to an increase of the number of non-zero elements
in ${\bf C}$ (see figure~\ref{fig:sparsity}). The orthonormalization is considered converged
when the absolute value of the largest off-diagonal element of ${\bf C^\dagger
S C}$ is lower than $\epsilon/100$.

The partial diagonalization of ${\bf H}$ does not need to be as precise as the orthonormalization,
so single precision is used to accelerate the calculation of ${\bf C^\dagger H C}$.
A threshold of $\epsilon$ is applied to the matrix elements of ${\bf H}$, and 
to the elements of ${\bf C}$ after a rotation of MOs. This larger cut-off increases the
sparsity of ${\bf C}$, but degrades the orthonormality of the MOs. To cure this latter problem,
an additional orthonormalization step among only the occupied MOs is performed after the
SCF has converged. 

%To be able to make relevant studies of large systems, one has to be able
%to compute accurately energy differences. As the size of the systems grow
%the total energies increase correspondingly, but the quantum chemist is still
%interested in energy differences in the order of 1~kcal/mol. The absolute error required
%is independent of the size of the system, but the relative error needs to be
%reduced when the size of the system grows to maintain the so-called {\em chemical
%accuracy}. Paradoxically, some additional approximations need to be introduced 
%to make computations feasible on large systems. To maintain the chemical accuracy
%these approximations need to be controlled.

\begin{table}
\caption{Total energies in a.u. obtained with the full diagonalization using
the Intel MKL implementation of the LAPACK library\citep{lapack} ($E_{\rm LAPACK}$) or the implementation proposed
in this work ($E$). $N$ is the number of water molecules.}
\label{tab:num_accuracy}
\centering
\begin{tabular}{|l|r|r|r|}
\hline
\multicolumn{1}{|c|}{$N$} &
\multicolumn{1}{c|}{$E_{\rm LAPACK}$} &
\multicolumn{1}{c|}{$E$} &
\multicolumn{1}{c|}{$\frac{E-E_{\rm LAPACK}}{E_{\rm LAPACK}}$} \\
\hline
\multicolumn{2}{|l|}{$\delta q_{\max} = 10^{-5}$} & &\\
\hline
184  &   -749.63938702 &   -749.63938696 &  8.0 10$^{-11}$ \\ %     4.486    3.832
368  &  -1499.33163492 &  -1499.33163479 &  8.7 10$^{-11}$ \\ %    28.517    8.056
736  &  -2998.74237166 &  -2998.74237136 &  1.0 10$^{-10}$ \\ %   194.465   19.989
1472 &  -5997.78267988 &  -5997.78267920 &  1.1 10$^{-10}$ \\ %  1398.725   53.071
2208 &  -8996.78192470 &  -8996.78192364 &  1.2 10$^{-10}$ \\ % 22710.115   91.273 #
3312 & -13495.30553928 & -13495.30553764 &  1.2 10$^{-10}$ \\ %  9636.091  149.783 # SV7
%4968 &                 & -20243.29421755 &   .  10$^{-10}$ \\ %  
\hline
\multicolumn{2}{|l|}{$\delta q_{\max} = 10^{-6}$} & &  \\
\hline
184  &   -749.63942605 &   -749.63942605 & $<$ 6.7 10$^{-12}$  \\ %     4.798    4.530
368  &  -1499.33168960 &  -1499.33168960 & $<$ 3.3 10$^{-12}$  \\ %    30.906   11.419
736  &  -2998.74251247 &  -2998.74251247 & $<$ 1.7 10$^{-12}$  \\ %   212.785   30.220
1472 &  -5997.78296964 &  -5997.78296962 &     3.3 10$^{-12}$ \\ %  1522.947   84.979
2208 &  -8996.78231327 &  -8996.78231325 &     2.2 10$^{-12}$ \\ % 25010.361   155.796 #
3312 & -13495.30617832 & -13495.30617828 &     3.0 10$^{-12}$ \\ %             258.950 # SV7
%4968 &                 & -20243.29519090 &      .  10$^{-12}$ \\ %             446.059
\hline
\end{tabular}
\end{table}

The total energies obtained with the divide and conquer LAPACK routine {\tt DSYGVD} are
compared to our implementation in table~\ref{tab:num_accuracy} with different
SCF convergence criteria. One can first remark that as the convergence
criterion of the SCF procedure gets lower, the error due to our implementation
decreases as $\epsilon$ decreases accordingly. Moreover, the difference in total energies
between the reference LAPACK result
and our implementation is orders of magnitudes below the error due to
the lack of convergence of the SCF procedure. The absolute error per water molecule
is below $2~10^{-10}$ using $\delta q_{\max} = 10^{-5}$, and around $3~10^{-12}$ for
$\delta q_{\max} = 10^{-6}$.

\subsection{Parallel efficiency and scaling}

In this section, the time values correspond to the timing of the {\em whole}
program, as obtained by the {\tt time} UNIX tool.

\begin{figure}
 \centering
 \includegraphics[angle=270,width=\columnwidth]{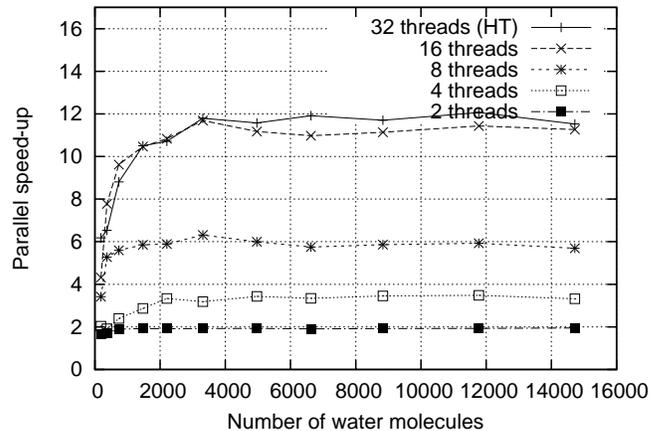}
 \caption{Parallel speedup as a function of the number of water molecules on the dual-socket server.
The data were renormalized to remove the effect of the turbo feature.
The 32 threads results were obtained using Hyperthreading (2 threads per core).}
 \label{fig:parallel_speedup_sv7}
\end{figure}

\begin{figure}
 \centering
 \includegraphics[angle=270,width=\columnwidth]{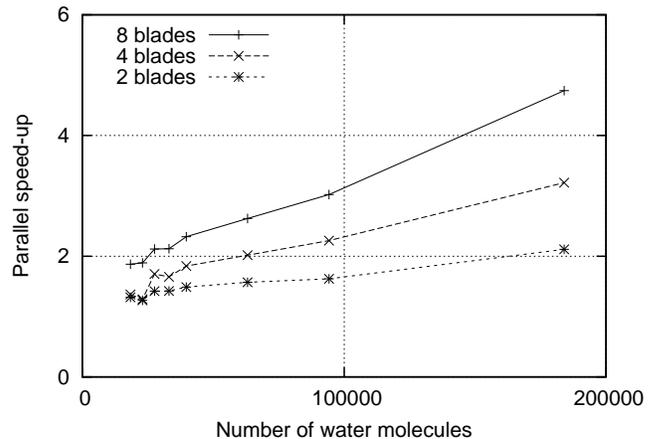}
 \caption{Parallel speedup as a function of the number of water molecules obtained on the Altix UV machine.
The reference is one 16-core blade.}
 \label{fig:parallel_speedup_uv}
\end{figure}

\begin{figure}
 \centering
 \includegraphics[angle=270,width=\columnwidth]{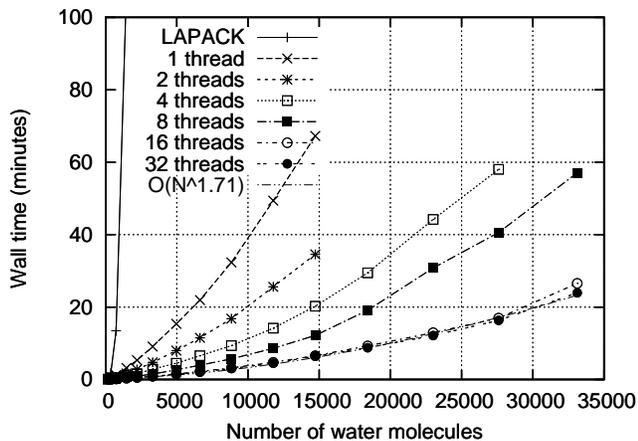}
 \caption{Total wall time (s) as a function of the number of water molecules obtained on the dual-socket server.
The 32 threads results were obtained using Hyperthreading (2 threads per
core).}
 \label{fig:wall_time_sv7}
\end{figure}

\begin{figure}
 \centering
 \includegraphics[angle=270,width=\columnwidth]{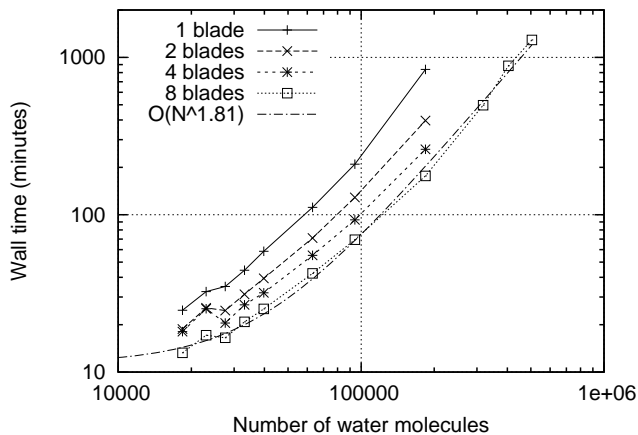}
 \caption{Total wall time (s) as a function of the number of water molecules obtained on the Altix UV machine.}
 \label{fig:wall_time_uv}
\end{figure}

Figure~\ref{fig:parallel_speedup_sv7} shows the parallel speed-up of the
program on the dual-socket server. When using all the 16 available physical
cores of the machine, the parallel efficiency converges quickly to value
between 11 and 12.
One can remark that Hyperthreading improves the parallel efficiency. Indeed,
as the memory latencies are the bottleneck, the CPU core can be kept busy for
one thread while another thread waits for data.
On the Altix UV (figure~\ref{fig:parallel_speedup_uv}), the speed-up is presented
as a function of the number of 16-core blades. Indeed, the important NUMA effects show up
when using more than one blade, since the inter-blade memory latency is significantly
higher than the intra-blade latency by a factor between 2 and 5. This explains
why the parallel efficiency is disappointing on the Altix UV: a speed-up close to 2
is obtained with 8 blades for small systems. However, when the systems is so large that
the total memory can't fit on a single blade, the 16-core reference suffers from a degradation
of performance due to the NUMA effects, so the comparison becomes more fair between the single-blade and the 8-blade
runs, and the parallel speed-up is much more acceptable, around 5.

From figures~\ref{fig:wall_time_sv7} and~\ref{fig:wall_time_uv}, the scaling of
the program as a function of the number of atoms is ${\cal O}(N^{1.7})$ on the
dual-socket server, and ${\cal O}(N^{1.8})$ on the Altix UV.
The global scaling is not linear. This is first due to the non-uniformity of
the memory access: the code is faster with smaller memory footprints since the
latencies are smaller. Secondly, in the expression of the Fock matrix
elements (Eqs.~\ref{H_dftb} and~\ref{H_dftb2}), for each atom one has to compute the Coulomb
contributions with all the other atoms. We chose not to truncate the $1/R$ function
in order to keep a good error control on the final energies, and this results
in a quadratic scaling with the number of atoms appearing for large systems
(10\% of the total time for 33~120 molecules on the dual-socket server, and
18\% for 504~896 molecules on the Altix UV).

\begin{figure}
 \centering
 \includegraphics[angle=270,width=\columnwidth]{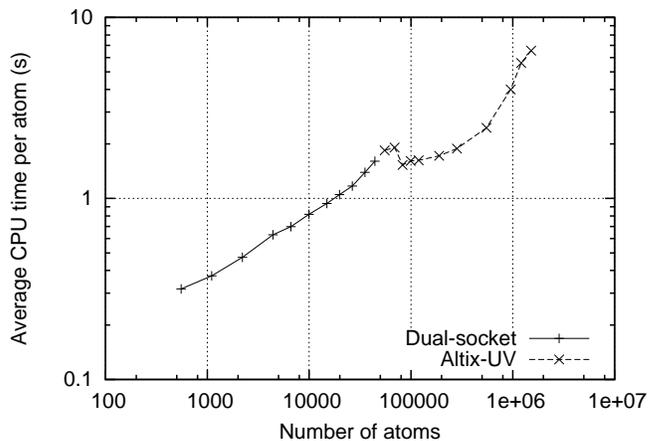}
 \caption{Average CPU time (s) per atom with increasing numbers of water molecules.}
 \label{fig:cpu_atom}
\end{figure}

The CPU time per atom is given in figure~\ref{fig:cpu_atom}. A similar benchmark was presented
with a linear-scaling approach in the density-matrix framework.\citep{cp2k} The authors
reported roughly $10$--$19$ CPU seconds per atom for systems with $50~10^3$--$1.1~10^6$ atoms.
It is difficult to compare quantitatively our timings with theirs, because
the method is different, molecular geometries are different and the hardware on
which the code ran is different. However, one can tell that our implementation
gives CPU timings in the same order of magnitude.
Another important point is that these authors used an MPI-based implementation, where they
report roughly 20\% of the time spent in MPI communication routines. Indeed, as linear-scaling
techniques are memory-latency bound, obtaining an efficient MPI implementation is a challenge
since MPI communications have a higher latency than memory accesses, even with low-latency network
hardware (Infiniband). Remark that our large simulations on the Altix UV are very efficient because
hardware prefetching occurs even when the target memory is located on a distant blade.
Therefore, the hardware makes automatically asynchronous inter-blade communication to further
reduce the memory latency bottleneck, which is not trivial to implement in a portable MPI framework.
This confirms also that our choice to defer the distributed parallelism to coarser
graining is a good choice.

\section{Summary and future work}

We have presented an efficient OpenMP implementation enabling MO-based SCC-DFTB
for very large molecular systems. As linear-scaling techniques reduce the
arithmetic intensity, these algorithms are inevitably memory-latency bound. Our
implementation was designed to minimize the effect of this bottleneck. We
have shown that we were able to run a calculation on more than 1.5 million
atoms (504~896 water molecules) with a very good efficiency on 128 cores of an
SGI Altix UV machine, a machine with very strong NUMA effects and memory
latencies that can be up to ten times larger than the memory latencies of a
standard desktop computer. 
The program is even more efficient for smaller systems (1~000--100~000 molecules),
the range which is the most suited to practical applications of DFTB. This aspect
will permit to run much longer molecular dynamics trajectories on medium-sized systems.
This OpenMP-only implementation makes it possible to run simulations with up to 100~000
atoms using standard servers, without the need of expensive low-latency network hardware.
The errors introduced by our approximations were shown to be below the error of the
SCF convergence threshold, which makes us trust our results for simulations with more than
a million of atoms, a domain where it is impossible to verify the validity of the result using an
exact diagonalization.
The next step of this project will be to implement distributed parallelism to
enable large scale Monte Carlo simulations of biological molecules in water.

%it would be now interesting to investigate the behavior of the code for chemically bond systems like for instance a large molecule. The creation of the 
%initial guess for moleular orbitals would then be less traightforward but several options could be followed in partitionung the system. In addition 
%it would be interesting to investigate how the spatial extension of the molecular orbitals evolve when compare to 
%pure water boxe. 

{\it Acknowledgments.}
The authors would like to thank ANR for support under Grant No ANR 2011
{{GASPARIM}}, ANR PARCS, This work was supported by ``Programme Investissements
d'Avenir'' under the program ANR-11-IDEX-0002-02 reference
ANR-10-LABX-0037-NEXT, CALMIP and Equipex Equip$@$Meso for computational
facilities.

\bibliography{sparse_dftb}
\bibliographystyle{unsrt}

\end{document}